\documentclass[aps]{revtex4}
\usepackage{amssymb}
\usepackage{amsmath}
\usepackage{graphicx}

\def \be {{\begin{equation}}}
\def \ee {{\end{equation}}}
\def \bea {\begin{eqnarray}}
\def \eea {\end{eqnarray}}

\begin{document}

\title{On electron-positron pair production using a two level on resonant
multiphoton approximation}
\author{I. Tsohantjis\renewcommand{\thefootnote}{*}\thanks{
Email: \texttt{ioannis2@otenet.gr}}, S. Moustaizis\renewcommand{%
\thefootnote}{**}\thanks{
Email: \texttt{moustaiz@science.tuc.gr}} and I. Ploumistakis%
\renewcommand{\thefootnote}{***}\thanks{
Email: \texttt{iploumistakis@isc.tuc.gr}}}
\affiliation{Technical University of Crete, Department of Sciences, Institute of Matter
Structure and Laser Physics,\\
Chania GR-73100, Crete, Greece}
\date{\today}

\begin{abstract}
We present an indepth investigation of certain aspects of the two level on
resonant multiphoton approximation to pair production from vacuum in the
presence of strong electromagnetic fields. Numerical computations strongly
suggest that a viable experimental verification of this approach using
modern optical laser technology can be achieved. It is shown that use of
higher harmonic within the presently available range of laser intensities
can lead to multiphoton processes offering up to 10$^{12}$ pairs per laser
shot. Finally the range of applicability of this approximation is examined
from the point of view of admissible values of electric field strength and
energy spectrum of the created pairs.
\end{abstract}

\maketitle




\section{Introduction}

Electron-positron pair production from vacuum in the presence of strong
electromagnetic fields is one of the most intriguing non-linear phenomena in
QED of outstanding importance specially nowadays where high intensity lasers
are available for experimental verification (for a concise review see \cite%
{greiner}, \cite{Fradkin}, \cite{Grib}). The theoretical treatment of this
phenomenon can be traced back to Klein \cite{Klein}, Sauter\cite{Sauter},
Heisenberg and Euler\cite{Heisenberg} but it was Schwinger \cite{Schwinger}
that first thoroughly examined this phenomenon, often called Schwinger
mechanism. Schwinger implementing the proper time method obtained the
conditions under which pair production is possible: the invariant quantities
$\mathcal{F=}\frac{1}{4}F_{\mu \nu }F^{\mu \nu }=-\frac{1}{2}\left( \mathcal{%
\vec{E}}^{2}-c^{2}\mathcal{\vec{B}}^{2}\right) $, $\mathcal{G=}\frac{1}{4}%
F_{\mu \nu }\tilde{F}^{\mu \nu }=c\mathcal{\vec{E}\cdot \vec{B}}$, where $%
F_{\mu \nu }$ and $\tilde{F}_{\mu \nu }=\frac{1}{2}\epsilon _{\mu \nu \alpha
\beta }$ $F^{\alpha \beta }$ are the electromagnetic field tensor and its
dual respectively, must be such that neither $\mathcal{F=}0$ , $\mathcal{G=}%
0 $ (case of plane wave field) nor $\mathcal{F>}0$ , $\mathcal{G=}0$ (pure
magnetic field). For the case of a static spacially uniform electric field
(where $\mathcal{F<}0$ , $\mathcal{G=}0$) he obtained a nonperturbative
result for the probability $w_{s}$ for a pair to be created per unit volume
and unit time to be $w_{s}(x)\sim \sum_{l=1}^{\infty }(1/l^{2})\exp (-\frac{%
l\pi m^{2}}{e\mathcal{E}})$. However in order to have sizable effects the
electric field strength $\mathcal{E}$ \ must exceed the \textit{critical
value} $\mathcal{E}_{c}$ $=\frac{mc^{2}}{e\lambda _{c}}\simeq 1.3\times
10^{18}V/m$ . Brezin and Itzykson \cite{Brezin} \ examined the case of \
pair creation in the presence of a pure oscillating electric field $\mathcal{%
E}$ (the presence of such electric field only can be achieved by using two
oppositely propagating laser beams so that in the antinodes of the standing
wave formed $\mathcal{F<}0$ and pair production can occur) by applying a
version of WKB approximation and treating the problem in an analogous way as
in the ionization of atoms(where the three basic mechanisms multiphoton,
tunneling and over the barrier ionization are present), considering the
pairs as bound in vacuum with binding energy $2mc^{2}$. The probability per
4-Compton volume of $\ e^{+}e^{-}$ pair creation is given by
\begin{equation}
w_{BI}=\frac{e^{2}\mathcal{E}^{2}}{\pi \hbar c}\frac{1}{g(\gamma )+\frac{%
\gamma g^{\prime }(\gamma )}{2}}\exp \left( -\frac{\pi m^{2}}{e\mathcal{E}}%
g(\gamma )\right) ,\;\gamma =\frac{mc\omega }{e\mathcal{E}}=\frac{\hbar
\omega \mathcal{E}_{c}}{mc^{2}\mathcal{E}}  \label{bi}
\end{equation}%
where $g(\gamma )=\frac{4}{\pi }\int_{0}^{1}\left( \frac{1-y^{2}}{1+\gamma
^{2}y^{2}}\right) ^{\frac{1}{2}}dy$ \ and the parameter $\gamma $ $=$%
(Photon\ energy/work\ of$\ \mathcal{E}\ $in\ a$\ \lambda _{Compton}$) is the
equivalent of the Keldysh parameter in the ionization of atoms. The formula
for $w_{BI}$ interpolates between two physically important regimes. For $%
\gamma \ll 1$ (high electric field strength and low frequency ), $g(\gamma
)=1-(1/8)\gamma ^{2}+O(\gamma ^{4})$, $w_{BI}$ $\sim \exp (-\pi (\mathcal{E}%
_{c}/\mathcal{E})g(\gamma ))$ and thus the adiabatic non-perturbative
tunneling mechanism dominates. When $\mathcal{E}$ $\ll \mathcal{E}_{c}$, $%
w_{s}(x)\simeq w_{BI}$ . For $\gamma \gg 1$ (low electric field strength and
high frequency), $g(\gamma )=(4/\pi \gamma )\ln (4\gamma /e)+O(1/\gamma
^{3}) $ and $w_{BI}$ $\sim (\mathcal{E}/\mathcal{E}_{c})^{2n_{0}}(1+O(1/%
\gamma ^{2}))$ ($n_{0}=2m/\omega $). This power-law behavior of $w_{BI}$ \
in the external field $\mathcal{E}$, is indicative of typical multiphoton
processes of order $n\geq $ $2m/\omega $ and $w_{BI}$ corresponds to the
n-th order perturbation theory in $\mathcal{E}$, n being the minimum number
of \ photons to create a pair. Soon after the work of \ Brezin and Itzykson,
in the work of Popov \cite{Popov1} (see also \cite{Nikishov2}, \cite{niki} ,%
\cite{Troup},\cite{popov2}) using the imaginary time method, the results of
\cite{Brezin} (and \cite{Schwinger}) were confirmed and investigated further
by determining also the pre exponential factor in $w_{BI}$ taking in to
account interference effects and treating again the system in analogous way
as in the ionization of atoms. In particular, with $\tau $ being the pulse
duration and $\lambda $ the electromagnetic wavelength, it was shown in \cite%
{Popov1} that for a spacially uniform oscillating electric field $\mathcal{E}
$ with frequency $\omega $ and under the conditions $\mathcal{E}$ $\ll
\mathcal{E}_{c}$, $\hbar \omega \ll $ $mc^{2}$ (which are both satisfied
from present laser technology) the probabilities over a Compton 4-volume $%
\lambda ^{3}\tau =\lambda ^{4}/c$ , can be obtained for any value of $\gamma
$ as a sum of probabilities $w_{n}$ of multiphoton processes of order $n$: $%
w_{P}=\sum_{n>n_{0}=2m/\omega }w_{n}$. For the exact rather lengthy formula
of $w_{n}$, which depends on $\gamma $, $g(\gamma )$ we refer the reader to
\cite{Popov1}, \cite{popov2}, \cite{Ringwald}. In the case $\gamma \ll 1$
the spectrum of $n\omega $ of the $n$-photon processes is practically
continuous giving the non-perturbative result $w_{P}\sim \left( \mathcal{E}/%
\mathcal{E}_{c}\right) ^{\frac{5}{2}}\exp (-\pi (\mathcal{E}_{c}/\mathcal{E}%
)g(\gamma ))$ (see \cite{popov2}). However in the typical multiphoton (and
of perturbative nature) case $\gamma \gg 1$, $w_{n}\sim \left( \mathcal{E}/%
\mathcal{E}_{c}\right) ^{2n}q\left( n-n_{0}\right) $ where $q\left(
n-n_{0}\right) =(1/2)e^{-2\left( n-n_{0}\right) }\int_{0}^{2\left(
n-n_{0}\right) }e^{t}t^{-1/2}dt$. The number of pairs created in the two
regimes are given by (see \cite{popov2})
\begin{eqnarray}
N(\tau ) &=&2^{-3/2}n_{0}^{4}\left( \mathcal{E}/\mathcal{E}_{c}\right) ^{%
\frac{5}{2}}\exp (-\frac{\pi \mathcal{E}_{c}}{\mathcal{E}}(1-\frac{1}{%
2\left( n_{0}\frac{\mathcal{E}}{\mathcal{E}_{c}}\right) ^{2}}))(\omega \tau
/2\pi ),\;\gamma \ll 1  \label{pop1} \\
N(\tau ) &\approx &2\pi n_{0}^{3/2}\left( \frac{8\mathcal{E}_{c}}{n_{0}e%
\mathcal{E}}\right) ^{-2n_{0}}(\omega \tau /2\pi ),\;\gamma \gg 1.
\label{pop2}
\end{eqnarray}%
One can easily see by comparing the above results that the multiphoton
processes are by far more efficient for pair production. Treatment of
Schwinger mechanism for non-oscillating electric fields and time dependent
magnetic fields see also \cite{WangWong}, \cite{gavgit} ,\cite{cal}, \cite%
{niki2}, \cite{kim}. For the role of temporal and spacial inhomogeneities in
the nonperturbative branch of pair production see \cite{kim}, \cite{dunne},
\cite{gies}, \cite{piazza}.

On the other hand the first experimental verification of $e^{-}e^{+}$ pair
production took place at SLAC ( E-144 experiment)\cite{Burke} where a
combination of nonlinear Compton scattering and multiphoton Breit-Wheeler
mechanism allowed for $e^{-}$ $e^{+}$ pair production to occur since the
available electric field intensities in the area of interaction of the
back-scattered photons with the laser used to produced them reached the
necessary values . The number of positrons measured in 21962 laser pulses
was 175$\pm 13$ and the multiphoton order of the process was found to be $%
n=5.1\pm 0.2(statistical)_{-0.8}^{+0.5}(systematic)$ , in very good
agreement with the theory. This experiment has led to a resent interest of
the subject especially as to whether modern laser technology can produce the
strong electric field required for experimental verification. As explicitly
analyzed by Ringwald \cite{Ringwald} both for the generalized WKB or
imaginary time methods, the optical laser technology available \cite{mourou}%
, as far as power densities and electric fields concerns, does not seem to
be implementable for experimental verification of $e^{-}e^{+}$ pair
creation, while for the X-Ray Free Electron Laser (XFEL) should be a very
promising facility (see also \cite{Melissinos}, \cite{chen}, \cite{tajima}).

However in a recent paper Avetissian et al \cite{Avetissian} treated the
problem of $e^{-}e^{+}$ production in a standing wave of oppositely directed
laser beams of plane transverse linearly polarized electromagnetic waves of
frequency $\omega $ and wavelength $\lambda $, using a two level multiphoton
on resonant approximation. As was shown there and qualitatively argued in
\cite{ctm} this approach if experimentally implemented will result in much
higher\ $e^{-}e^{+}$ production rate for the case of conventional
femto-second lasers systems. The main difference of this approach to the one
mentioned above is the resonance condition. Also, since the fundamental
parameter of the theory is $\xi =e\mathcal{E}\mathbf{/}mc\omega \leq 1$, the
results of this method can only be compared with the corresponding ones from
the perturbative multiphoton regime $\gamma \geq 1$ above.

The aim of this article is to investigate further this approximation mainly
focusing on numerical computations that convincingly support the possibility
of experimentally detectable pair creation with available optical laser
technology. Of special interest is the use of higher harmonics such as 3$%
\omega $ and 5$\omega $. Moreover the close resemblance of this
approximation with multiphoton ionization of \ atoms highlights a lot of \
the physically interesting characteristics that one might expect to detect
in the laboratory. In particular, ultrashort laser systems such as Nd-Yag or
Ti-Sapphire, with an intensity at the fundamental frequency $\omega $, of \
the order of 10$^{22}W/m^{2}$ , when working on the multiphoton on resonant
regime, is shown to produce number of pairs of the order of 10$^{8}$ or more
per laser shot. On the other hand such laser systems, with intensities up to
the order of 10$^{30}W/m^{2}$ , can provide higher harmonics pair creation,
such as 3$\omega $ and 5$\omega $, where the number of pairs is shown to
reach up to 10$^{12}$ per laser shot. As is demonstrated one can keep the
frequency fixed and gradually change the electric field strength, and
perform that for each frequency chosen. However for the laser systems under
consideration it is difficult to adjust $\mathcal{E}$ while being on
resonant and moreover there are limitations on the increase of it as will be
shown . What is experimentally viable is to increase the frequency and,
without having to focus in the diffraction limit, increase the intensity so
that the resulting increase in $\mathcal{E}$ will be such that the ratio $%
\xi =e\mathcal{E}\mathbf{/}mc\omega $ \ is fixed. In section two we briefly
present the results of \cite{Avetissian} referring the reader to that
article for their derivation. In section three we investigate the behavior
of the probability density and the number of pair created by the fundamental
and higher harmonics of a conventional laser with respect to changes in the
electric field strength and the energy spectrum of the created
electrons(positron). We end this section by showing that there exist bounds
on the values of the electric field strength, the multiphoton order and the
energy spectrum for the two level on resonant multiphoton approximation to
hold. Finally in section four we conclude with suggested ways of
experimental verification and future line of research. All numerical results
have been produce for an Nd-Yag laser of photon energy $1.17eV$ and
intensity 1.35$\times 10^{22}W/m^{2}$ and using Mathematica and Maple
packages.

\section{Basic results of the two-level on resonant multiphoton
approximation of pair production from vacuum.}

Following \cite{Avetissian} a standing wave $\overrightarrow{A}=2%
\overrightarrow{A}_{0}\cos \overrightarrow{k}\overrightarrow{r}\cos \omega t$
is formed by two oppositely propagating laser beams of frequency $\omega $
and wavelength $\lambda $ (see also \cite{Ringwald}). Pair production
essentially occurs close to the antinodes and in spacial dimensions $l$ $\ll
\lambda $ so that $\overrightarrow{k}\overrightarrow{r}=\frac{2\pi }{\lambda
}l$ \ \ is very small and thus the spacial dependence of the resulting wave
can be disregarded, that is $\overrightarrow{A}=2\overrightarrow{A}_{0}\cos
\omega t$. Moreover since the interaction Hamiltonian is of the form $%
\overrightarrow{p}\overrightarrow{A}$ the most significant contribution in
the pair creation process in the regions of antinodes will be at the
direction along the electric field. Due to space homogeneity in these
regions the 4-momentum of a particle is conserved, transitions occur between
two energy levels from $\mathcal{-}E$ to $E$ by the absorption of $n$
photons and the multiphoton probabilities will have maximum values for
resonant transitions
\begin{equation}
n=2E/\omega  \label{res1}
\end{equation}%
Non-linear solutions of the Dirac equation under these conditions were
obtained resulting to the following probability for an n-photon $e^{-}e^{+}$
pair creation, summed over the spin states
\begin{equation}
W_{n}=2f_{n}^{2}\frac{\sin ^{2}\left( \Omega _{n}\tau \right) }{\Omega
_{n}^{2}}  \label{wn1arm}
\end{equation}%
where
\begin{equation}
f_{n}=\frac{E}{4p\cos \theta }\left( 1-\frac{p^{2}\cos ^{2}\theta }{E^{2}}%
\right) ^{\frac{1}{2}}n\omega J_{n}(4\xi \frac{mp\cos \theta }{E\omega })
\label{fn}
\end{equation}%
$\xi $ is the relativistic invariant parameter given by,
\begin{equation}
\xi =\frac{e\mid \mathcal{E}_{o}\mid }{mc\omega }\lesssim 1  \label{ksi}
\end{equation}%
$\Omega _{n}$ and $\Delta _{n}$ is the 'Rabi frequency' of the Dirac vacuum
at the interaction with a periodic electromagnetic field and respectively
given by,
\begin{equation}
\Omega _{n}=\sqrt{f_{n}^{2}+\frac{\Delta _{n}^{2}}{4}}\ll \omega ,\;
\label{rabi}
\end{equation}%
$\theta $\ is the angle between the momentum of $e^{-}\left( e^{+}\right) $
and $A_{0}$, $\mathcal{E}_{o}$ is the amplitude of the electric filed\
strength of one incident wave, $\Delta _{n}=2E-n\omega $ is the detuning of
resonance, and $\tau $ is the interaction time. In obtaining the above
probability it has been assumed without loss of generality that $p_{z}=0$
since there is a symmetry with respect to the direction of $A_{0}$ (taken to
be the Oy axis) and thus $\mathbf{p}=(p_{x}=p\sin \theta ,p_{y}=p\cos \theta
,0)$. As usual in applying the resonance approximation on a two level system
the probability amplitudes are slow varying functions which is equivalently
expressed here by the condition in (\ref{rabi}), corresponding to such field
intensities for which the condition in (\ref{ksi}) is satisfied. For short
interaction time i.e. when $\Omega _{n}\tau \ll 1$, $\frac{\sin ^{2}\left(
\Omega _{n}\tau \right) }{\Omega _{n}^{2}}\rightarrow 2\pi \tau \delta
\left( \Delta _{n}\right) $ and the differential probability per unit time
summed over the spin states in the phase-space volume $Vd^{3}p/\left( 2\pi
\right) ^{3}$ \ is $dw_{n}=\frac{1}{2\pi ^{2}}f_{n}^{2}\delta (2E-n\omega
)Vd^{3}p$ which after integration over the $e^{-}$($e^{+}$) energy, the
angular distribution of a n-photon differential probability of the created $%
e^{-}$, $e^{+}$ pair, per unit time in unit space volume ($V=1$), on exact
resonance is given by:
\begin{equation}
\frac{dw_{n}}{do}=\frac{n\omega }{8\pi ^{2}}f_{n}^{2}\left( n^{2}\omega
^{2}-4m^{2}\right) ^{\frac{1}{2}}  \label{dwndoarm}
\end{equation}%
where $do=\sin \theta d\theta d\varphi $. The total angular distribution of
probability is $\frac{dw}{do}=\sum_{n=n_{0}}\frac{dw_{n}}{do}$\ (where $%
n_{0}=2mc^{2}/\hbar \omega $ is the threshold number of photons for the pair
production process to occur) and integrating over the solid angle we obtain
the total probability per unit time in unit space volume of the $e^{-}$, $%
e^{+}$ pair production $w=\sum_{n=n_{0}}w_{n}$ as:
\begin{eqnarray}
w &=&\sum_{n=n_{0}}\frac{n^{5}\omega ^{5}}{32\pi p}((\frac{2Z_{0}^{2}}{%
4n^{2}-1}-1)J_{n}^{2}\left( Z_{0}\right) +\frac{Z_{0}^{2}J_{n-1}^{2}\left(
Z_{0}\right) }{2n(2n-1)}+\frac{Z_{0}^{2}J_{n+1}^{2}\left( Z_{0}\right) }{%
2n(2n+1)}  \notag \\
&&-\frac{4p^{2}}{n^{2}\omega ^{2}}\frac{Z_{0}^{2n}}{\left( 2n+1\right)
\left( n!\right) ^{2}2^{2n}}\times \;_{2}F_{3}\left( n+\frac{1}{2},n+\frac{1%
}{2};n+1,2n+1,n+\frac{3}{2};-Z_{0}^{2}\right) )  \label{warm}
\end{eqnarray}%
where $Z_{0}=\left( \frac{4\xi m}{\omega }\right) \left( 1-\frac{4m^{2}}{%
n^{2}\omega ^{2}}\right) ^{\frac{1}{2}}$. The total number of pairs $N$
created for a given laser characteristics can be estimated by (see \cite%
{Avetissian})
\begin{equation}
N\sim wV\tau ,\;V\sim \sigma ^{2}l  \label{npairs}
\end{equation}%
where $V$ is the space-volume, $\sigma $ is the cross section radius, $l\ll
\lambda $ as stated above and $\tau $ is the interaction time. For focused
optical lasers in the diffraction limit $\sigma \sim \lambda $ $\sim
10^{-6}m $ and $\tau \sim 10^{-14}s$. For the investigation that will follow
\begin{equation}
\frac{dN_{n}}{do}=\frac{dw_{n}}{do}V\tau  \label{dn}
\end{equation}%
is the angular distribution of the number of pairs created from an $n$%
-photon process and
\begin{equation*}
N_{n}=w_{n}V\tau
\end{equation*}%
is the number of pairs created from that process.

\section{Numerics and applicability of the on resonant multiphoton
approximation of pair production from vacuum.}

As can be seen from section II a basic role in the physical interpretation
of the numerical computations that will follow, is played by the function $%
f_{n}$ (Rabi frequency on exact resonance), as the probabilities and number
of produced pairs obtained are heavily depend on its behavior (see (\ref%
{wn1arm}), (\ref{dwndoarm})). For a given value of the $\xi $ and $n$, as
can be seen from (\ref{fn}), $\ f_{n}$ and all derived angular dependent
quantities in the above section, maximizes at $\ \theta =0$ and this is true
for every $\xi $ and $n$. Consequentially we shall concentrate our analysis
at this angle of observation of created pairs. Not only this simplifies the
numerics that will be presented below but also helps to clarify the behavior
of this approximation in particular as far as future experimental
verification. From now on $c$ and $\hbar $ should be explicitly stated in
the formulas. On exact resonance, $n$ is given by (see \ref{res1})
\begin{equation}
n=2E/\hbar \omega =2qmc^{2}/\hbar \omega ,\;q\geq 1,  \label{res}
\end{equation}%
where we have expressed the energy $E$ of the created electron (positron) in
terms of its rest energy as $E=qmc^{2}$. Thus $q$ characterizes the spectrum
of the created pairs. At $\ \theta =0$, a suitable expression for $f_{n}$, $%
f_{n}$, can be obtained from (\ref{fn}) with $E=qmc^{2}$, $p=(1/c)\sqrt{%
E^{2}-m^{2}c^{4}}$, and using the asymptotic behavior of the Bessel
function\ $J_{n}(x)$ at $x\simeq n$ (see also \cite{Avetissian}). In fact,
as can be seen from (\ref{res}) for optical lasers where $\omega $ is very
small (of the order of $eV$), $n$ is very large and as $\xi \lesssim 1$, the
argument of the Bessel function in (\ref{fn}), which now becomes $x=\frac{%
2n\xi }{q}\left( 1-\frac{1}{q^{2}}\right) ^{\frac{1}{2}}$, is also very
large and of the same order as $n$, not mentioning Bessel's extreme
sensitivity on $\xi $ too. Thus to obtain executable numerical computations,
we shall from now\ on adopt this asymptotic behavior of the Bessel function
by writing $J_{n}(x)=J_{n}(nsecha)=(1/\sqrt{2\pi ntanha})\exp \left(
ntanha-na\right) $ where $a$=$sech^{-1}(\frac{2\xi }{q}\left( 1-\frac{1}{%
q^{2}}\right) ^{\frac{1}{2}})$. Then $f_{n}$ is given by
\begin{equation}
f_{n}=\frac{1}{4}\left( q^{2}-1\right) ^{-\frac{1}{2}}n\hbar \omega \frac{%
\exp \left( ntanha-na\right) }{\sqrt{2\pi ntanha}}  \label{fna}
\end{equation}%
The function $f_{n}$ can now be used together with (\ref{dwndoarm}) and (\ref%
{dn}), to obtain the number of pairs at $\ \theta =0$, $N_{0}=\frac{dN_{n}}{%
do}|_{\theta =0}$ as
\begin{equation}
N_{0}=\frac{dN_{n}}{do}|_{\theta =0}=\frac{1}{4\pi ^{2}}\frac{V\tau }{V_{e}}%
\frac{q\sqrt{q^{2}-1}}{m^{2}c^{4}}f_{n}^{2}  \label{numberpairs}
\end{equation}%
where $V_{e}=7.4\times 10^{-59}m^{3}s$ is the four Compton volume of an
electron.

\begin{figure}[tbp]
\includegraphics[width=17.5cm]{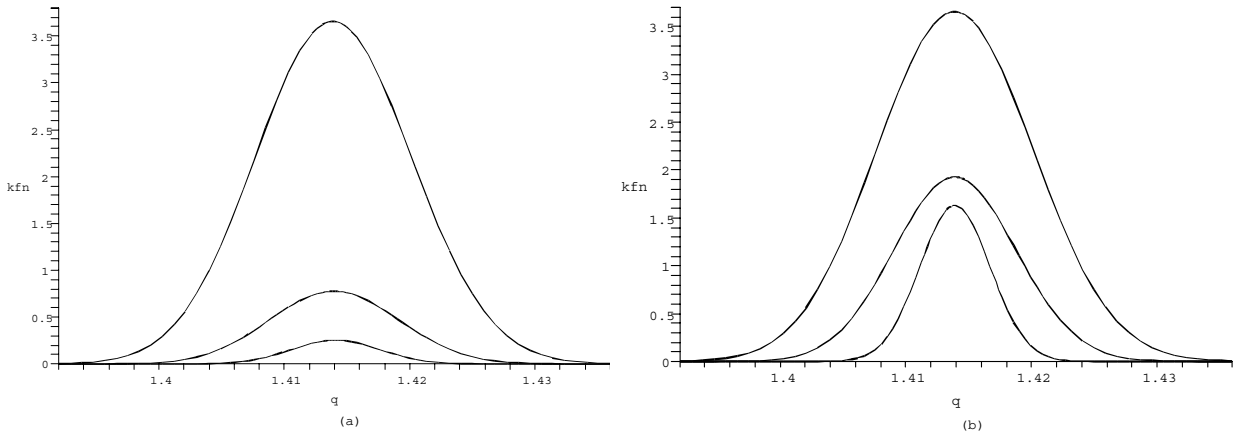}\newline
\caption{(a) The envelops of $f_{n}$(at $\protect\theta =0$), as a function
of the units of rest energy $q$, for $\protect\xi =0.9987$ and $5\protect%
\omega $ (top curve), $\protect\xi =0.9990$ and $3\protect\omega $ (middle
curve) and $\protect\xi =0.9995$ and $\protect\omega =1.17eV$ (bottom
curve), $k=10^{21}$. (b) The envelops of $f_{n}$(at $\protect\theta =0$) as
a function of the units of rest energy $q$ for $\protect\xi =0.9987$ and for
$\protect\omega =1.17eV$ (bottom curve with $k=10^{40}$), $3\protect\omega $
(middle curve with $k=10^{24}$), $5\protect\omega $ (top curve with $%
k=10^{21}$).}
\label{fig1}
\end{figure}

Using (\ref{res}), (\ref{fna}),\ the envelope of $\ f_{n}$ as a function of $%
q$ can be plotted for fixed values of \ $\xi $. This allow to investigate
the envelop of \ $\ f_{n}$, from electric field strength , frequency of
radiation or both point of view. In fig.\ref{fig1}(a) (see also \cite%
{Avetissian}), \ we plot the envelops of $\ \ f_{n}$, for the case of $%
\omega =1.17eV$ , $3\omega $ and $5\omega $ and for values of $\xi =0.9995$,
$0.9990$ and $0.9987$ respectively. The corresponding electric fields $%
\mathcal{E}_{o}$ are approximately given by (\ref{ksi}) as $3.\,\allowbreak
024\,2\times 10^{12}$ V/m, $9.\,\allowbreak 068\,1\times 10^{12}V/m$, $%
1.\,\allowbreak 510\,9\times 10^{13}V/m$. Each point in a curve of fig.\ref%
{fig1}(a) corresponds via (\ref{res}) to an order $n$ multiphoton process
and to an energy $E=qmc^{2}$ of the \ electron (positron) to be created in
the area of antinodes under the application of fixed field strength and
frequency. The most probable process corresponds to the peaks of the curves
which will be labeled with the triplet ($n_{p}$ ,$q_{p}$, $\xi $). For the
three cases of fig.\ref{fig1}(a) , using common differential calculus, we
find peaks approximately at ($1.2369\times 10^{6}$, $1.41408$, $0.9995$), ($%
4.1226\times 10^{5}$, $1.41395$, $0.9990$) and ($2.4734\times 10^{5}$, $%
1.41387$, $0.9987$) respectively.

A quite interesting case when dealing with higher harmonics is to
investigate the behavior of $f_{n}$(at $\ \theta =0$) for $\xi $ fixed. As
we change from $\omega $ to $2\omega $, $3\omega $ etc., an appropriate,
experimentally viable, increase of the laser intensity can lead $\mathcal{E}%
_{o}$ to increase by the same amount as $\omega $. In fig.\ref{fig1}(b),
such case is presented for $\xi =0.9987$ and $\omega =1.17eV$ , $3\omega $
and $5\omega $ where the corresponding envelops have peaks ($n_{p}$ ,$q_{p}$%
) at (1.2367$\times 10^{6}$, 1.41390), (4.1223$\times 10^{5}$, 1.41388) and
(2.4734$\times 10^{5}$, 1.41387) respectively. Both from fig.\ref{fig1}(a,
b), it is seen that passing to higher harmonics the peak value of $\ f_{n}$
increases rapidly leading to an increase of the probability of pairs
created., with a subsequent decrease of \ the most probable multiphoton
order $n_{p}$ and corresponding energy $E_{p}=$ $q_{p}mc^{2}$ of
electron(positron) created. Moreover the range of the energy spectrum of the
pairs broadens thus facilitating their observation: from approximately $%
0.720MeV$ to $0.726MeV$ which is for $\omega $ , to, $0.715MeV$ to $0.731MeV$
which is for 5$\omega $. An explanation for the choices of values for $\xi $
will be conferred till the end of this section.

\begin{figure}[tbp]
\includegraphics[width=17.5cm]{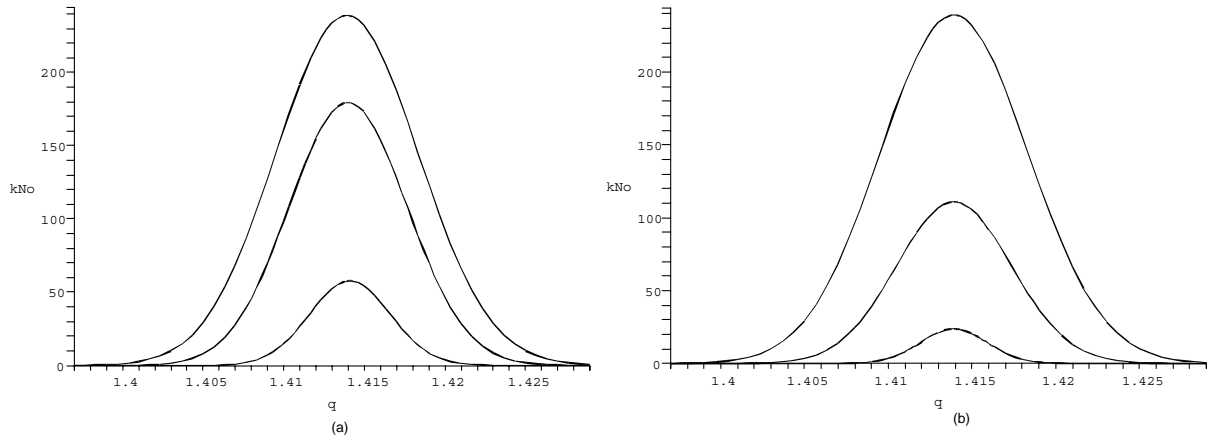}\newline
\caption{(a) Envelop of number of pairs created $N_{0}$, as a function of
the units of rest mass $q$, at angle $\protect\theta =0$ for the multiphoton
processes $\protect\omega =1.17eV$ (bottom curve with $k=10^{-7}$), $3%
\protect\omega $ (middle curve with $k=10^{-7}$) and $5\protect\omega $ (top
curve with $k=10^{-8}$) of fig.1(a).(b)Envelop of number of pairs created $%
N_{0}$, as a function of the units of rest mass $q$, at angle $\protect%
\theta =0$ and $\protect\xi =0.9987$, for the multiphoton processes $\protect%
\omega =1.17eV$ (bottom curve with $k=10^{29}$), $3\protect\omega $ (middle
curve with $k=10^{-2}$) and $5\protect\omega $ (top curve with $k=10^{-8}$)
of fig.1(b).}
\label{fig2}
\end{figure}

Corresponding to each envelop of $\ f_{n}$ we can plot the envelop of \ the
number of pairs created by n-photon processes\ $N_{0}$ $,$ as a function of
\ $q$, using (\ref{dwndoarm}), (\ref{dn}), (\ref{res}), (\ref{fna}), (\ref%
{numberpairs}). Examples are presented in fig.\ref{fig2}(a) (see also fig.%
\ref{fig1}(a)) for the cases $\omega =1.17eV$ , $3\omega $, $5\omega $ and
for values of $\xi =0.9995$, $0.9990$ and $0.9987$ respectively. The four
volume used in each case has been calculated by (\ref{npairs}), with $\tau
\sim 10^{-14}s$ , $\lambda =1.074\times 10^{-6}$m and $\sigma \sim 10^{-5}m$%
, leading to $V\tau \sim \sigma ^{2}l\tau \sim \sigma ^{2}(0.1\lambda
/k)\tau ,$ where $k=1,3,5$ for the corresponding harmonics. Note that we do
not necessarily have to work in the diffraction limit $\sigma \sim \lambda $
as the number of pairs created is adequately high for observation, while to
conform with the developed approximation where $l\ll \lambda ,$ the choice $%
l=0.1\lambda /k$ demonstrates the fact that when going to higher harmonics
the area close to the antinodes that the pair creation essentially happens
decreases. Each of these curves essentially give the energy spectrum of the
created number of pairs at $\theta =0$ after the application of a fixed
electric field strength and laser frequency and for all $n$-photon process
at exact resonance. Their peaks can be labeled by the triplet ($N_{p}$ ,$%
q_{p}$, $\xi $), $N_{p}$ being the maximum (and most probable) number of
pairs created for the $n_{p}$-photon processes of fig.\ref{fig1}(a). These
three cases have peaks approximately at (5.856$\times 10^{8}$,1.41408, $%
0.9995$), (1.815$\times 10^{9}$,1.41395, $0.9990$) and ( 2.372 $\times
10^{10}$,1.41387, $0.9987$) respectively. The corresponding values of \ $%
\mathcal{E}_{o}$ and the range of the energy spectrum are as those in fig.%
\ref{fig1}(a) above. Experimentally such curves are important as one can
detect the electron(positron) energies coming up from the various $n$-photon
processes for a given $\mathcal{E}_{o}$ and laser frequency and compare with
these theoretical estimates.

The case corresponding to fig.\ref{fig1}(b) is presented in fig.\ref{fig2}%
(b), where for $\omega =1.17eV$ , $3\omega $ and $5\omega $ and \ $\xi
=0.9987$ fixed (and thus for $\mathcal{E}_{0}$, 3$\mathcal{E}_{0}$ and $5%
\mathcal{E}_{0}$), the corresponding envelops have peaks ($N_{p}$ ,$q_{p}$)
approximately at (2.430$\times 10^{-28}$, 1.41390), (1.104$\times 10^{4}$,
1.41388) and ( 2.391 $\times 10^{10}$, 1.41387) \ corresponding to the $%
n_{p} $-photon processes of fig.\ref{fig1}(b). It is easily seen from both
these figures that going to higher harmonics, the number of pairs increases
very rapidly with simultaneous increase of the range of energies of the
pairs but decrease of their maximum energy.

We turn now to a commonly experimentally verifiable behavior of multiphoton
processes given by the log-log plot of the number of particles created
versus the value of electric field strength $\mathcal{E}_{o}$. In fig.\ref%
{fig3} we present the log-plots of the number of pairs $N_{0}$ as a function
of $\xi $ , using (\ref{dwndoarm}), (\ref{dn}), (\ref{res}), (\ref{fna}), (%
\ref{numberpairs}), for three on resonant multiphoton process with $%
n_{1}\sim 1.\,\allowbreak 233\times 10^{6}$ ($q\sim 1.41$), $n_{2}\sim
1.237\times 10^{6}$ ($q\sim 1.4141$) and $n_{3}\sim 1.\,\allowbreak
242\times 10^{6}$ ($q\sim 1.42$) chosen from the bottom curve of fig.\ref%
{fig1}(a) where $\omega =1.17eV$ is kept fixed (see also bottom curve of fig.%
\ref{fig2}(a)). Note that the energies of the created particles for each of
the above on resonance multiphoton processes are close enough given
approximately by $E_{1}$ $\sim $0 .721 MeV, $E_{2}$ $\sim $ $0.723\,$MeV and
$E_{3}$ $\sim \allowbreak 0.726$ MeV respectively while the range of change
of $\mathcal{E}_{o}$ producing observationally enough pairs is between $%
3.\,\allowbreak 023\,8\times 10^{12}$ V/m to $3.\,\allowbreak 024\,5\times
10^{12}$V/m. The range of change of $\mathcal{E}_{o}$ (and thus of $\xi )$
is very small even for higher harmonics because of the extreme sensitivity
of the Bessel function and its approximation in $\xi $. This suggests that
an experimental verification of such curves is rather difficult for optical
lasers. As $\omega $ is fixed and thus the appearance of the different on
resonant multiphoton processes originate only from the different energies
involved (see values of $q$), crossings in these curves, which traditionally
appear in multiphoton ionization, are not to be expected. Further more, as
will be explained in the end of this section, such curves terminate from
above for a maximum value of $\mathcal{E}_{o}$ (and thus of $\xi )$ .

\begin{figure}[tbp]
\includegraphics[width=17.5cm]{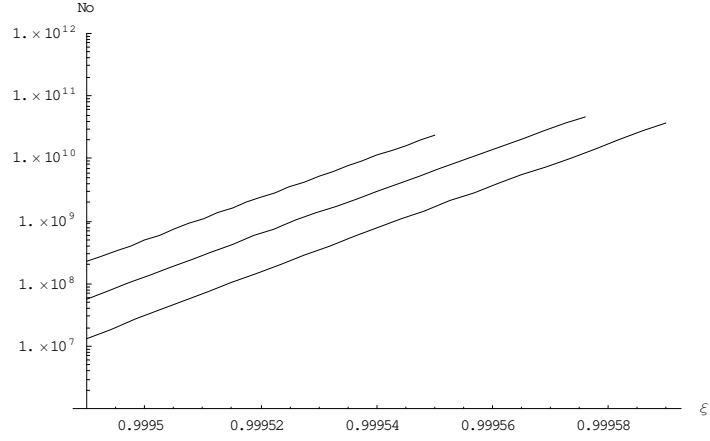}\newline
\caption{Log-plot of the number of pairs created $N_{0}$, as a function of $%
\protect\xi $, at angle $\protect\theta =0$, for three multiphoton processes
from the bottom curve of fig.1, with $q=1.41$ (middle curve) , $q\sim
\protect\sqrt{2}$ (top curve) and $q=1.42$ (bottom curve).}
\label{fig3}
\end{figure}

\begin{figure}[tbp]
\includegraphics[width=18.5cm]{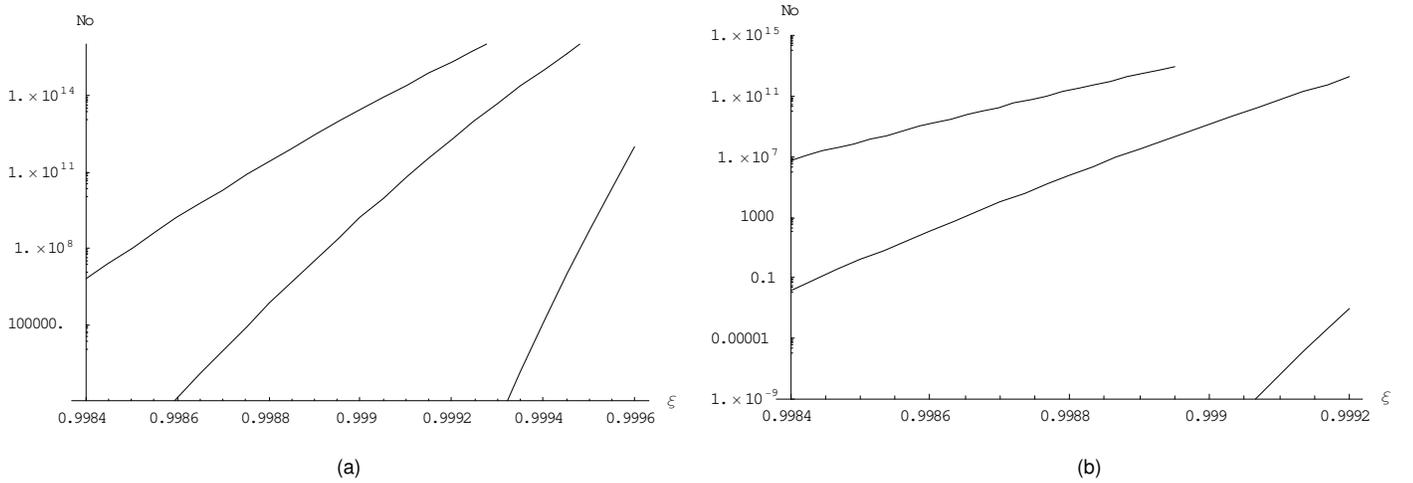}\newline
\caption{(a) Log-plot of the number of pairs created $N_{0}$, as a function
of $\protect\xi $, for the most probable multiphoton processes of fig.1(a)
with $\protect\omega =1.17eV$ (bottom curve) , $3\protect\omega $ (middle
curve) and $5\protect\omega $ (top curve). (b) Log-plot of the number of
pairs created $N_{0}$, as a function of $\protect\xi $, for the most
probable multiphoton processes of fig.1(b) with $\protect\omega =1.17eV$
(bottom curve) , $3\protect\omega $ (middle curve) and $5\protect\omega $
(top curve).}
\label{fig4}
\end{figure}

In fig.\ref{fig4}(a) we give the log-plot of the number of pairs $N_{0}$
versus $\xi $ for the most probable multiphoton processes of $\omega =1.17eV$
, $3\omega $, $5\omega $ of fig.\ref{fig1}(a) (see also fig.\ref{fig2}(a))
where ($n_{p}$ ,$q_{p}$, $\xi $)$\sim $($1.2369\times 10^{6}$, $1.41408$, $%
0.9995$), ($4.1226$ $\times 10^{5}$, $1.41395$, $0.9990$) and ($2.4734$$%
\times 10^{5}$, $1.41387$, $0.9987$) respectively . In contrast with the
case presented in fig.\ref{fig3}, crossings are expected as the laser
frequency changes. However for the developed approximation, the values of $%
\xi $ where these occur are not applicable as $\xi $ $>1$. Similar results
arise when we consider the most probable multiphoton processes ($n_{p}$ ,$%
q_{p}$, $0.9987$) of fig.\ref{fig1}(b) (see also fig.\ref{fig2}(b)) and are
presented in fig.\ref{fig4}(b), where for $\omega $, 3$\omega $ and 5$\omega
,$ ($n_{p}$ ,$q_{p}$)$\sim $(1.2367$\times 10^{6}$, 1.41390), (4.1223$\times
10^{5}$, 1.41388) and (2.4734$\times 10^{5}$, 1.41387) respectively.

\begin{figure}[tbp]
\includegraphics[width=16.5cm]{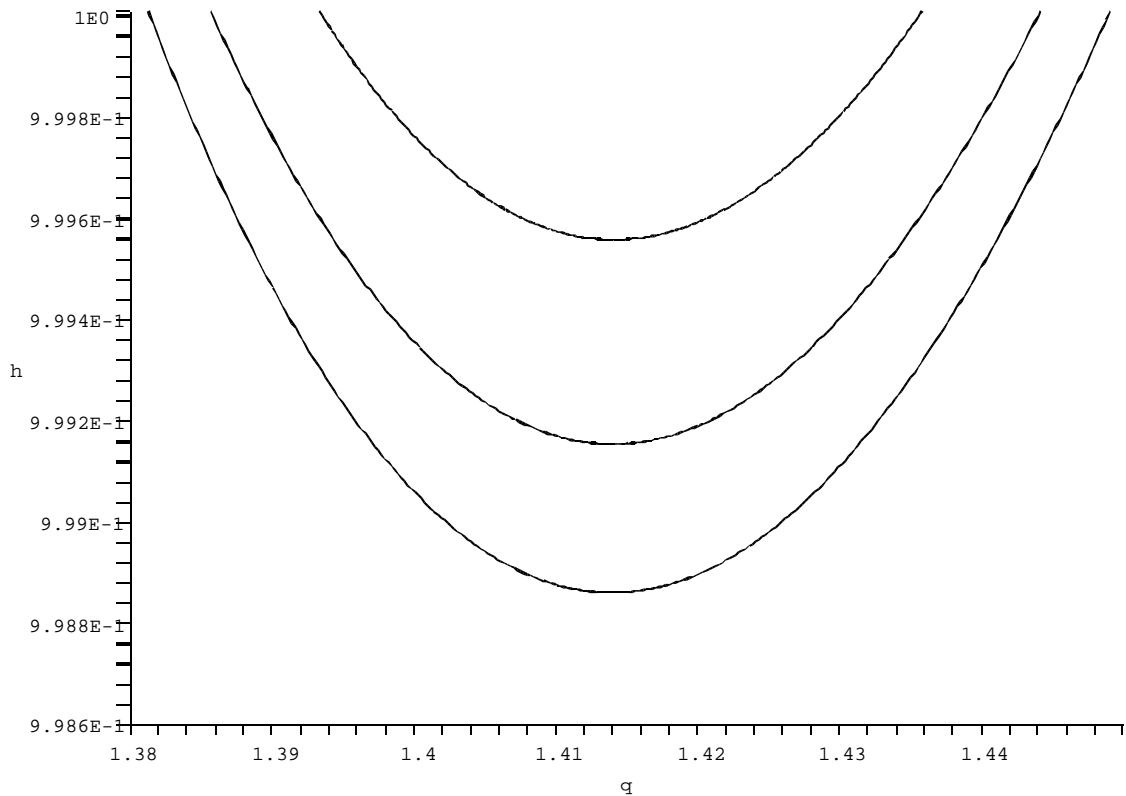}\newline
\caption{Upper bound $h$ of $\protect\xi $ as a function of $q$ for the
cases $\protect\omega =1.17eV$ (top curve), $3\protect\omega $ (middle
curve) and $5\protect\omega $ (bottom curve).}
\label{fig5}
\end{figure}

Given an initial laser frequency and power density, the obvious question to
be raised concerns on one hand the range of possible multiphoton processes
that can be obtain within this approximation (or equivalently the range of
energy of the created pairs per rest energy of e$^{-}$ , $q$ ) and on the
other hand the range of values of $\xi $ (or equivalently of the electric
field strength $\mathcal{E}_{0}$) for which these are realized. The physical
acceptable values of $\xi $, $q$ have not only to conform with the condition
of applicability of resonant approximation $\Omega _{n}\ll \omega $ (i.e.$%
\xi \precsim 1$) but also to energy considerations stating that the energy
per laser shot, $E_{b}$ , provided by the incident beam , should not be less
than the total energy of the pairs created, that is
\begin{equation}
E_{b}\geqslant 2qmc^{2}N  \label{encon}
\end{equation}%
where $N$ is the total number of pairs created. $E_{b}$ can be calculated
from the available power density of the laser $S_{b}=\frac{1}{\mu _{0}c}$ $%
\mathcal{E}_{0}^{2}$ as
\begin{equation}
E_{b}=S_{b}\pi \sigma ^{2}\tau  \label{laserenergy}
\end{equation}%
where $\sigma $ is the radius of the cross section and $\tau $ is the pulse
duration. To get a sufficiently convincing answer to the above question we
can consider the energy difference
\begin{equation}
\Delta E_{b}=S_{b}\pi \sigma ^{2}\tau -2qmc^{2}N_{0}  \label{energydif}
\end{equation}%
which by means of (\ref{fna}) and (\ref{numberpairs}) is considered as a
function of $\xi $(or $\mathcal{E}_{0}$) and $q$(or $n$) . Keeping $E_{b}$
fixed (i.e. for given laser characteristics $\omega $, $S_{b}$, $\sigma $, $%
\tau $) and for a given $q\geq 1$, $\xi $ can be increased up to a value $%
\xi =h$ (or maximum $\mathcal{E}_{0}$) for which $\Delta E_{b}=0$(minimum
physically acceptable value of $\Delta E_{b}$) provided that $h\ngtr 1$.
Consequentially, for given values of $q,$ we can quit sufficiently estimate
the applicability of the present approximation by numerically computing the
upper bounds $h$ of $\xi $, using $S_{b}\pi \sigma ^{2}\tau =2qmc^{2}N_{0}$
(of course we could also keep $\xi \precsim 1$ fixed and numerically compute
$q$, but for experimental reasons, we are merely interested in the maximum
applicable $\mathcal{E}_{0}$ for the present approximation to hold). In fig.%
\ref{fig5}

we plot the maximum admissible values $h$ of $\xi $ (or $\mathcal{E}_{0}$)
as a function of $q$ (and thus of $n$), for the three cases $\omega $, 3$%
\omega $ and 5$\omega $ where computations have been performed using $\Delta
E_{b}=0$ for $\omega =1.17eV$, $S_{b}=s\times 1.35\times 10^{22}W/m^{2}$ ($%
s=1,3^{2},5^{2}$ respectively), $\sigma \sim 10^{-5}m$ and $\tau \sim
10^{-14}s$. The\ factor $s$ in $S_{b}$ is justified by the approach adopted
to increase the laser intensity in order to increase $E_{b}$ , rather than
going to the diffraction limit ($\sigma \sim \lambda ^{\prime }$) to
increase it, as this would be experimentally tedious when going to higher
harmonics $\omega ^{\prime }=k\omega $, where $\lambda ^{\prime }=\lambda /k,
$ $k=1,2,3...$. From the curves of fig.\ref{fig5} the range of the
applicable on resonant multiphoton processes can easily be read off via the
range of values of $q$ shown and using (\ref{res}). Moreover the maximum
applicable values of $\xi $ (and thus via (\ref{ksi}) of $\mathcal{E}_{0}$)
for each one of them can also be read off. Points ($q$, $h$) for $h>1$ are
unacceptable for the two level on resonant approximation of pair production.
Also because of the existence of $h$ for each $q$ (and $n$) points in the
log-plots of figs \ref{fig3},\ref{fig4}(a, b), where $\xi >h$ should be
disregarded, and thus the curves for these plots should be terminated at $%
\xi =h$ or equivalently at $\mathcal{E}_{0}=\mathcal{E}_{0\max }=hmc\omega
^{\prime }/e$. That is also why crossing points cannot be present in the
log-plots.

As an example of the above consider the three peak points of the curves $%
\omega $, $3\omega $, $5\omega $ in fig.\ref{fig2}(a). The $q_{p}$ values of
these points are situated close to the bottom of the corresponding curves of
fig. \ref{fig5} from which we can infer their corresponding $h$s to be
approximately $h\sim 0.99956$, $0.99916$, $0.99886$. Moreover as can be seen
from fig. \ref{fig4}(a, b), when $\xi $ approaches $h$ the number of pairs
created for the corresponding $n_{p}$ multiphoton processes reaches a
maximum value. This explains the choices of $\xi $ chosen in the above
numerical computations to be close to $h$. Consequentially points ($\xi $, $%
N_{0}$) in figs \ref{fig3}, \ref{fig4}(a, b) with values of $\xi >h$ \emph{%
should not} be taken in to account.

Another important consequence of the upper bound $h$, concerns the value of $%
\xi $ chosen when examining the spectrum of created pairs, for fixed $\omega
^{\prime }$, via plots of fig.\ref{fig2}(a, b). For simplicity consider $%
\omega ^{\prime }=\omega $. In fig.\ref{fig3} the three terminal points of
these curves, which maximize $N_{0}$, corresponds to the points (1.41,
0.99957), ($\sqrt{2}$, 0.99956), (1.42, 0.99959) of the $\omega $-curve of
fig.\ref{fig5}, ($\sqrt{2}$, 0.99956) being the lowest point of it. If one
chooses to work with an $h\neq $0.99956, say $h=$0.99959, then fig.\ref{fig3}
shows that energies with $q<1.42$ can never be observed. However plots such
as fig.\ref{fig2}(a) with $\xi =$0.99959 can be drawn showing that points
with values of $q$ in the physically forbidden range do contribute in $N_{0}$%
. Obviously this is a completely unphysical situation and should be taken
care in experimental verification of plots such as fig.\ref{fig2}(a, b). In
fact the only consistent value of $\xi $ is the one of the lowest point ($%
q_{l\text{ }}$, $\xi =h_{l}$) of the $\omega ^{\prime }$-curve of fig.\ref%
{fig5} as this guarantees both observability of all energies around $q_{l%
\text{ }}=q_{p\text{ }}$ as given in fig.\ref{fig2}(a, b) and maximization
of $N_{0}$ for this $q_{p}$.

\section{Conclusion}

From the above analysis\ it is evident that present ultrashort laser
technology seems to suffices in order to experimentally verify the validity
of $e^{+}e^{-}$ pair production from vacuum using a two level on resonance
multiphoton approximation. In particular, emphasis has been given in the
implementation of higher harmonics such as 3$\omega $ and 5$\omega $ while
the electric field strengths required, are obtained by increasing the laser
energy rather than focusing to the diffraction limit. This improves the
model in various advantageous ways. The need of higher harmonics is dictated
by the limitation imposed by the upper value of electric field $\mathcal{E}%
_{o}$ of the fundamental due to the condition $\xi =\frac{e\mathcal{E}_{o}}{%
mc\omega }\lesssim 1$. In order to work with $\xi \lesssim 1$ but increase
the $\mathcal{E}_{o}$ higher $\omega $ values are necessarily.

Firstly, as shown in figs \ref{fig1}, \ref{fig2}, the range of the created
spectrum widens and the maximum number of pairs created increases
drastically reaching $N_{0}=$10$^{12}$ pairs per laser shot for 5$\omega $
while, because of the resonant condition, the electric fields needed are low
$\mathcal{E}_{0}$ $\sim 10^{13}V/m$, compared with other multiphoton
approximations such as the one leading to (\ref{pop2}). In fact this is
mainly why there is no need to focus in the diffraction limit to achieve
such electric fields as present laser energies and achievable power can
provide them.

Secondly the confirmation of the power law behavior of the number of
pairs created as a function of electric field strength, typical of
multiphoton processes, is demonstrated by figs \ref{fig3},
\ref{fig4}, showing again a drastic increase of $N_{0}$ in higher
harmonics. However such log-plots can not probably be subjected to
experimental verification since the range of \ change of \
$\mathcal{E}_{0}$ is very small and thus difficult if not
technically impossible to be performed. However what it is suggested
in the
present work is the verification of higher harmonic curves of fig.\ref{fig2}%
, of the number of pairs $N_{0}$versus their spectrum, when measuring the
number and the momenta of the created electrons(positrons) at angle $\theta
=0$.

Finally the range of applicability of this approximation have been
investigated and the results are presented in fig.\ref{fig5}. In particular
working with a chosen frequency, for each q there exists a maximum value $%
\xi =h$ and thus a maximum electric field $\mathcal{E}_{0\max }$ that can be
used. As has been demonstrated by the analysis of fig.\ref{fig5} in section
III there important consequences for a potential experimental verification
of the suggested plots of fig.\ref{fig2}(a, b). Consequently one can
describe the following attractive experimental scenario. Initially one
should choose a laser energy $E_{b}$ capable of generating a higher harmonic
$\omega ^{\prime }=k\omega $ beam. Then by appropriate focusing, increase
the electric field at the value $\mathcal{E}_{0\max }=h_{l}mck\omega /e$
where $h_{l}$ is the lowest value of the $k\omega $ curve of fig.\ref{fig5},
and form the standing wave as required by the theory. The number of pairs $%
N_{0}$ created at the antinodes versus their spectrum will be given
by figures such as those of fig.2(a, b) drawn for $\xi =h_{l}$. Then
$N_{0}$ maximizes for pairs with energy $E=2q_{p}mc^{2}$ where
($q_{p}$, $h_{l}$) is the lowest point of the $k\omega $ curve of
fig.\ref{fig5}. Higher harmonics thus give a wider pair spectrum and
a lower $\mathcal{E}_{0\max }$ value required, both been of great
experimental advantage.

In concluding one should state that use of XFEL technology (equivalent to
ultrahigh harmonics) overcomes the difficulties of so high order of
multiphoton processes present in the optical regime, while giving a wider
range of electric field changes. Investigations along the lines of the
present article of the application of the resonant approximation using XFEL
are in progress.


\begin{thebibliography}{99}
\bibitem{greiner} W. Greiner, B. Muller, J. Rafelski, `Quantum
Electrodynamics of Strong Fields', Springer -- Verlag, Berlin, 1985.

\bibitem{Fradkin} E. S. Fradkin, D. M. Gitman and Sh. M. Shvartsman,
'Quantum Electrodynamics with unstable vacuum' Springer-Verlag, Berlin, 1991.

\bibitem{Grib} A. A. Grib, S. G. Mamaev and V. M. Mostapanenko, 'Vacuum
Quantum Effects in Strong Fields' Atomizdat, Moscow, 1998; Fr iedmann
Laboratory Publishing, St. Petersburg 1994.

\bibitem{Klein} O. Klein, Z. Phys., \textbf{53}, 157 (1929).

\bibitem{Sauter} F. Sauter, Z. Phys. \textbf{69}, 742 (1931).

\bibitem{Heisenberg} W. Heisenberg, H. Euler, Z. Phys. \textbf{98}, 718
(1936).

\bibitem{Schwinger} J. W. Schwinger, Phys. Rev., \textbf{82}, 664 (1951).

\bibitem{Brezin} E. Brezin and C. Itzykson, Phys. Rev. D \textbf{2}, 1191
(1970).

\bibitem{Popov1} V.S. Popov, JETP Lett. \textbf{13}, 185 (1971); Sov. Phys.
JETP \textbf{34}, 709 (1972); Sov. Phys. JETP \textbf{35}, 659 (1972); V.S.
Popov and M. S. Marinov, Sov. J. Nucl. Phys.\textbf{16}, 449 (1973) ; JETP
Lett. \textbf{18}, 255 (1974); Sov. J. Nucl. Phys., \textbf{19}, 584 (1974).

\bibitem{Nikishov2} A. I. Nikishov, Nucl. Phys. B\textbf{21}, 346 (1970).

\bibitem{niki} N.B. Narozhnyi and A. I. Nikishov, Sov. J. Nucl. Phys.\textbf{%
11}, 596 (1970); \ Sov. Phys. JETP, \textbf{38}, 427 (1974).

\bibitem{Troup} G.J. Troup and H.S. Perlman, Phys. Rev. D \textbf{6}, 2299
(1972).

\bibitem{popov2} V.S. Popov, Phys. Let. A\textbf{298}, 83 (2002).

\bibitem{Ringwald} A. Ringwald, Phys. Let. B \textbf{510}, 107 (2001).

\bibitem{WangWong} R.C. Wang and C.Y. Wong, Phys. Rev. D \textbf{38}, 348
(1988).

\bibitem{gavgit} S. P. Gavrilov and D. M. Gitman Phys. Rev. D \textbf{53},
7162 (1995).

\bibitem{cal} G. Calucci, "Pair production in a time dependent magnetic
field", hep-th/9905013.

\bibitem{niki2} A. I. Nikishov, "On the theory of scalar pair production by
a potential barrier", hep-th/0111137.

\bibitem{kim} S. P. Kim and Don N. Page, Phys. Rev. D\textbf{73} : 065020,
(2006); "Schwinger pair production in electric and magnetic fields"
hep-th/0301132.

\bibitem{dunne} G. V. Dunne, Q. Wang, H. Gies and C. Schubert, Phys. Rev. D%
\textbf{73} : 065028, (2006); hep-th/0602176.

\bibitem{gies} H. Gies and K. Klingmuller, Phys. Rev. D\textbf{72} : 065001,
(2005); hep-ph/0505099.

\bibitem{piazza} A. DiPiazza, Phys. Rev. D\textbf{70} : 053013, (2004);

\bibitem{Burke} D.L. Burke et. al., Phys. Rev. Let., \textbf{79}, 1626
(1997).

\bibitem{mourou} M. Perry and G. Mourou, Science \textbf{264}, 917 (1994).

\bibitem{Melissinos} A.C. Melissinos, in Quantum Aspects of Beam Physics,
Proc.15th Advanced ICFA Beam Dynamics Workshop, Monterey, Cal., 4-9 Jan 1998
(World Scientific, Singapore, 1998) p. 564.

\bibitem{chen} P. Chen and C. Pellegrini, in Quantum Aspects of Beam
Physics, Proc.15th Advanced ICFA Beam Dynamics Workshop, Monterey, Cal., 4-9
Jan 1998 (World Scientific, Singapore, 1998) p. 571.

\bibitem{tajima} P. Chen and T. Tajima, Phys. Rev. Lett. \textbf{83}, 256
(1999).

\bibitem{Avetissian} H. K. Avetissian, A. K. Avetissian, G. F. Mkrtchian and
Kh. V. Sedrakian, Phys. Rev. E \textbf{66}, 016502 (2002).

\bibitem{ctm} C. Kaberidis, I. Tsohantjis and S. Moustaizis 'Multiphoton
approach on pair production under the light of recent experimental and
theoretical investigations', Proceedings of the Sixth International
Symposium `Frontiers of Foundamental and Computational Physics' Udine,
Italy, 26-29 September 2004, Sidharth B.G, Honsell F., de Angelis A. (Eds.)
2005 pp. 279-283
\end{thebibliography}
\end{document}